\documentclass[twocolumn,superscriptaddress,amsmath,amssymb,floatfix]{revtex4}

\usepackage{graphicx}
\usepackage{dcolumn}
\usepackage{bm}

\graphicspath{{Figures/}}
\setkeys{Gin}{width=\linewidth}

\begin{document}

\title{Coulomb Gap in Graphene Nanoribbons}

\author{S. Dr\"oscher}
\affiliation{Solid State Physics Laboratory, ETH Zurich, 8093 Zurich, Switzerland}
\author{H. Knowles}
\affiliation{Solid State Physics Laboratory, ETH Zurich, 8093 Zurich, Switzerland}
\author{Y. Meir}
\affiliation{Physics Department, Ben Gurion University, Beer Sheva 84105, Israel}
\author{K. Ensslin}
\affiliation{Solid State Physics Laboratory, ETH Zurich, 8093 Zurich, Switzerland}
\author{T. Ihn}
\affiliation{Solid State Physics Laboratory, ETH Zurich, 8093 Zurich, Switzerland}

\begin{abstract}
We investigate the density and temperature-dependent conductance of graphene nanoribbons with varying aspect ratio. Transport is dominated by a chain of quantum dots forming spontaneously due to disorder. Depending on ribbon length, electron density, and temperature, single or multiple quantum dots dominate the conductance. Between conductance resonances cotunneling transport at the lowest temperatures turns into activated transport at higher temperatures. The density-dependent activation energy resembles the Coulomb gap in a quantitative manner. Individual resonances show signatures of multi-level transport in some regimes, and stochastic Coulomb blockade in others.
\end{abstract}

\maketitle

Monolayer graphene shows impressive material stability, even if shaped into nanostructures of about 10 nm in size \cite{Ponomarenko08,Wang08,Ihn10}. Its electronic properties are tunable by gate electrodes \cite{Novoselov04,Molitor07} like conventional semiconductors, while its conductivity competes with that of metals. Graphene nanoribbons have the potential to be used in nanoelectronics \cite{Wang08}, and graphene nanoconstrictions are the basic building blocks for quantum devices \cite{Ihn10}.

The transport properties of graphene ribbons and constrictions on a SiO$_2$ substrate have been one of the puzzles for the understanding of graphene nanostructures. Early predictions of an energy gap in ribbons \cite{Nakada96,Peres06,Gunlycke07} have triggered intense experimental \cite{Han07,Chen07,Todd09,Stampfer09,Molitor09,Liu09,Han10,Gallagher10,Oostinga10,Terres11,Tombros11} and theoretical research \cite{Gunlycke07,Areshkin07,Sols07,Querlioz08,Lherbier08,Adam08,Evaldsson08,Mucciolo09,Schubert09,Martin09,Ihnatsenka09,Klos09,Libisch11}. It has become evident experimentally that localized states due to edge and bulk disorder suppress the conduction and lead to a transport gap \cite{Todd09,Stampfer09,Molitor09} rather than a true band gap. In addition, experiments indicate the formation of an interaction driven Coulomb gap \cite{Todd09,Stampfer09,Molitor09,Liu09,Han10}. A wealth of theoretical ideas ranging from Anderson localization \cite{Lherbier08,Adam08,Evaldsson08,Mucciolo09,Schubert09} to Coulomb blockade \cite{Sols07,Martin09} try to explain the phenomenology.

We show in this Letter that electronic transport in narrow nanoribbons is dominated by a chain of one or multiple quantum dots forming due to disorder. Not only is the conductance activated between conductance resonances, but the activation energy at each density corresponds to the Coulomb gap. At the lowest temperatures, cotunneling is present. Our experiments indicate that transport through graphene nanoribbons can be understood based on the mesoscopic details of the sample in a single-particle picture including Coulomb blockade. In contrast to recent suggestions \cite{Han10,Oostinga10}, there is no indication that additional energy scales or mechanisms are necessary to describe the observed behavior.

\begin{figure}[ht]
  \begin{center}
    \includegraphics{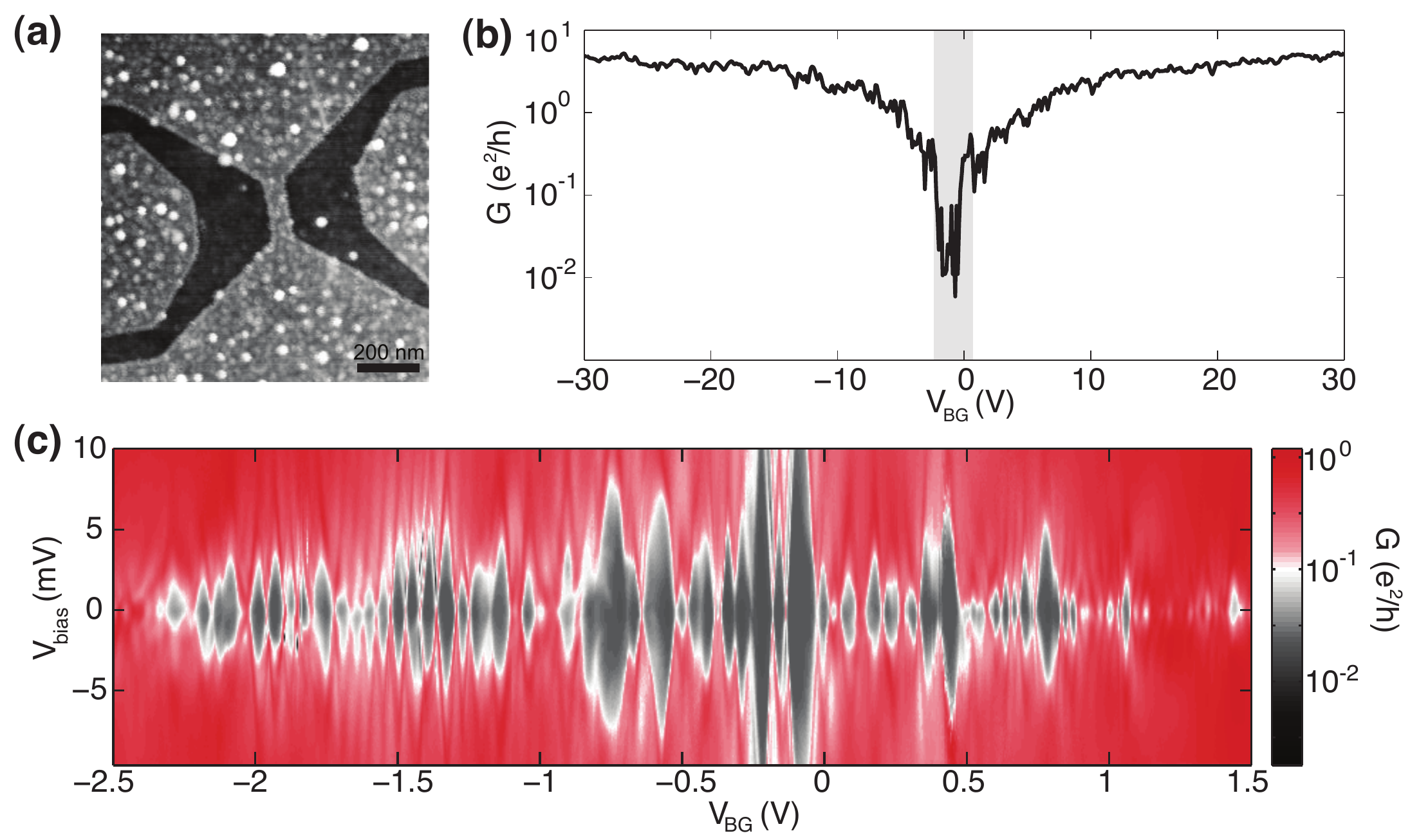}
    \caption{(a) Scanning force micrograph of the nanoribbon investigated here ($L$ = 200 nm, $W$ = 75 nm). (b) $G(V_\mathrm{BG}$) in a large density range showing the transport gap around $V_\mathrm{BG}$ = -2 V. The measurement was taken at $T$ = 1.25 K with a source-drain bias of $V_\mathrm{bias}$ = 500 $\mu$V. (c) Finite-bias measurement inside the transport gap (same temperature as (b)).} 
    \label{fig1}
  \end{center}
\end{figure}

Graphene nanoribbons with widths below 120 nm and lengths of 100 nm and 200 nm were fabricated as described in Ref. \onlinecite{Molitor09} on a SiO${_2}$ layer covering the highly doped Si-substrate which serves as a global back-gate. Five different devices [length (nm) $\times$ width (nm): 200$\times$75, 100$\times$45, 100$\times$80, 100$\times$100, 100$\times$120] were characterized in detail within this study, all showing the same qualitative behavior.

The measurements were carried out in the variable temperature insert of a $^4$He cryostat with a base temperature of 1.25 K. The conductance was measured using standard lock-in techniques at 13 Hz.

In the following we limit the detailed presentation of the results to the representative device displayed in the scanning force micrograph in Fig. \ref{fig1}(a) with ribbon length $L$ = 200 nm and width $W$ = 75 nm. Changing the back-gate voltage from hole transport to electron transport allows us to locate the charge neutrality point to be around -2 V in back-gate voltage [see Fig. \ref{fig1}(b)]. Like in earlier studies \cite{Han07,Chen07,Todd09,Stampfer09,Molitor09,Liu09,Han10,Gallagher10,Oostinga10,Terres11,Tombros11}, a region of suppressed conductance is present around this gate voltage [shaded region in Fig. \ref{fig1}(b)]. This regime is commonly referred to as the transport gap and gives an estimate for the amplitude of the potential inhomogeneity in the ribbon \cite{Todd09,Stampfer09,Molitor09}.

The behavior of the conductance in Fig. \ref{fig1}(b) for $V_\mathrm{BG}$ $>$ -2 V is qualitatively similar to earlier observations in narrow disordered channels in Si-inversion layers \cite{Fowler82}, where the large conductance fluctuations at low charge carrier densities were attributed to structure in the density of states leading to hopping transport between strongly localized states. They are smeared out as either the temperature or the charge carrier density is increased. Inside the transport gap the small value of the conductance $G$ $\ll$ $e^2/h$ indicates that the system is strongly localized \cite{Thouless77}. In the investigated device the size of the gap is $\Delta V_\mathrm{BG}$ $\approx$ 3.5 V in good agreement with the statistics from other measurements and the scaling law introduced by Han et al. \cite{Han07}, which relates the width of the nanoribbon with the energy of the transport gap. The studies in Refs. \onlinecite{Gallagher10} and \onlinecite{Terres11} have shown that the transport gap is largely independent of the ribbon length.

\begin{figure}[ht]
  \begin{center}
    \includegraphics{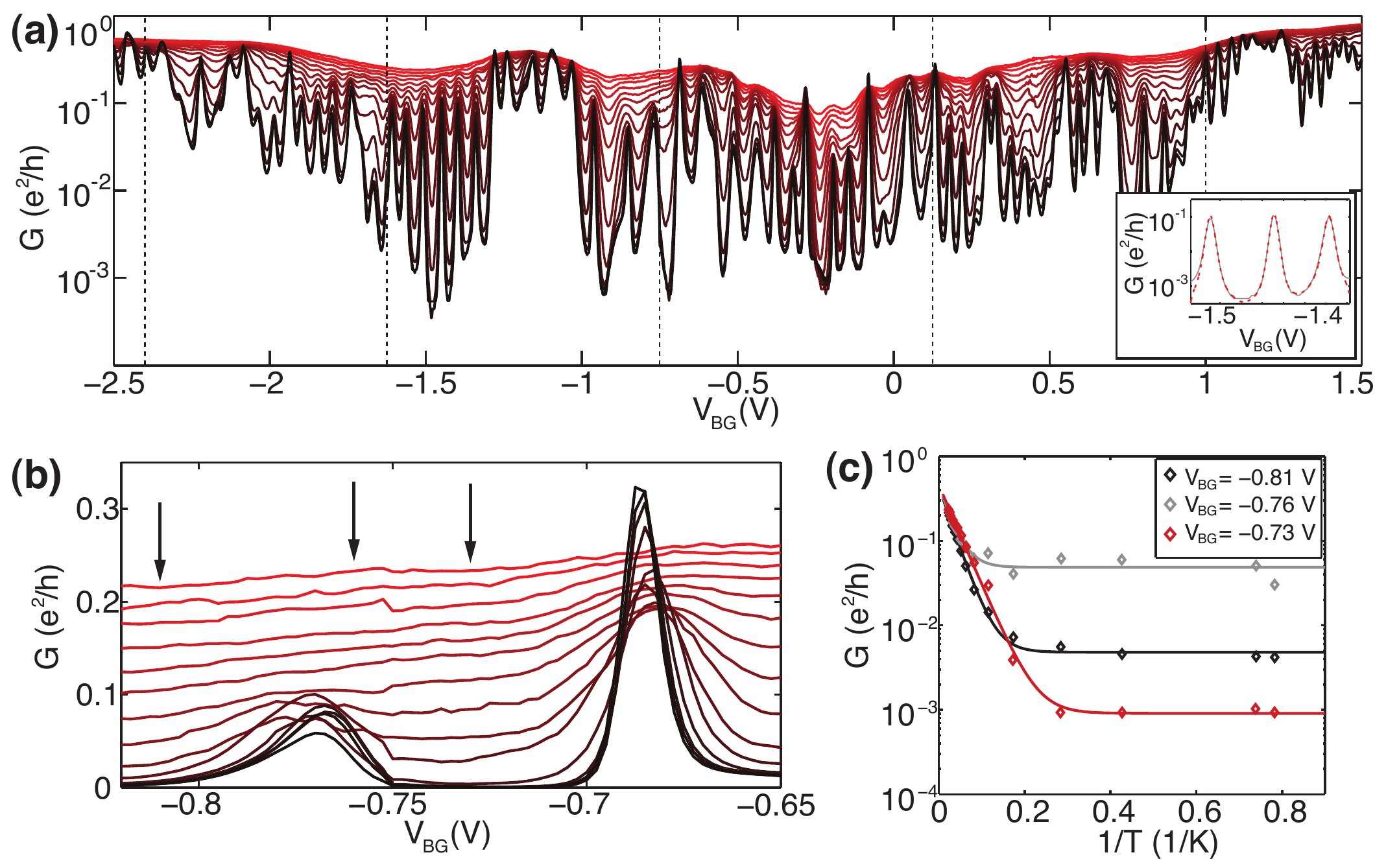}
    \caption{(a) $T$-dependence of $G$ inside the transport gap at $V_\mathrm{bias}$ = 100 $\mu$V. Different curves are taken at T = 1.25 K to 45 K (black to red lines). Inset: Coulomb resonances (grey line) reconstructed by a convolution of three Lorentzians with the derivative of the Fermi distribution (red dotted line). (b) Zoom into two exemplary peaks of (a). The left peak is broadened and grows with increasing $T$ and the right peak exhibits an overall decrease of $G$ with temperature. (c) $G$ as a function of 1/$T$ at three positions in $V_\mathrm{BG}$ indicated by arrows in (b). Solid lines are fits to the data according to Eq. (\ref{G}).}
    \label{fig2}
  \end{center}
\end{figure}

For further characterization of the device we measured the conductance inside the transport gap at finite-biases applied between source and drain. The recorded diamonds of suppressed current are shown in Fig. \ref{fig1}(c). Diamonds of different sizes can be identified which sometimes overlap. However, in some regimes (e.g. around $V_\mathrm{BG}$ = -1.5 V) resonances at zero source-drain bias are observable which separate adjacent diamonds from each other indicating single quantum dot behavior rather than transport through multiple dots. This phenomenology is usually referred to as stochastic Coulomb blockade \cite{Kemerink94}.

Coulomb interactions play an important role in graphene and lead to the formation of a Coulomb gap \cite{Han10}. Following the approach by Molitor et al. \cite{Molitor09}, a measure for the spatial extent of the localized islands in a device can be obtained by finite-bias spectroscopy. In devices of different widths, the extracted charging energies $E_\mathrm{c}$  of the largest diamond in the gap are inversely proportional to the width of the nanoribbons and only very weakly dependent on the length \cite{Gallagher10,Terres11}. Again, the device investigated in our work falls well into the statistics of the data published earlier with $E_\mathrm{c}$ = 5-10 meV.

Several microscopic pictures have been introduced to explain the formation of a transport gap in graphene nanoribbons. Lattice defects at the edges could cause Anderson localization \cite{Evaldsson08, Mucciolo09} which would suppress transmission around the charge neutrality point. An alternative picture suggests the formation of quantum dots along the ribbon due to potential fluctuations \cite{Sols07}. A small confinement gap is required in the latter case to prevent Klein tunneling between the puddles. Experimental transport data could so far be interpreted in both models. Knowledge about transport mechanisms, which we investigate here in thermal activation studies, may help to understand where and how localization comes about.

Fig. \ref{fig2}(a) displays the back-gate voltage dependent conductance at various temperatures for the complete transport gap. To obtain this temperature dependence of $G$, the investigated back-gate voltage range was split into intervals of about 1 V as indicated in Fig. \ref{fig2}(a) by the vertical dashed lines. In these sections $G$ was measured at stepwise increasing temperatures between 1.25 and 45 K. In all sections it was verified that the low-temperature Coulomb peak spectra were identical before and after the thermal cycle. A number of approximately 100 Coulomb resonances are visible in the region of suppressed conductance. For any minimum between two resonances the conductance increases for increasing temperatures. Even at the highest temperatures the conductance approaches but does not exceed $e^2/h$ meaning that the system remains in the strongly localized regime.

If we associate with each conductance resonance the addition of a single electron to the system, the transport gap corresponds to a density of states of $\approx$5$\times$10$^{16}$ m$^{-2}$eV$^{-1}$. This value is in good agreement with Ref. \onlinecite{Droescher10}, where the density of states was determined from the quantum capacitance of a top-gated large-area device.

Fig. \ref{fig2}(b) shows a close-up for Coulomb resonances with distinctly different behaviors. The amplitude of the left peak grows with $T$ and the peak broadens at the same time until it is finally swamped away by the rising background. The signature of the right peak is a maximum peak value of $G$ at the lowest temperature which drops to a local minimum at intermediate temperatures and recovers as $T$ is increased further. Such a behavior is found only for those $\approx$10 $\%$ of the resonances in the investigated back-gate window, which are particularly sharp at low temperatures.

For a single quantum dot this observation has been explained by the interplay between temperature, the single-particle level spacing $\Delta_s$, and the coupling of the energy levels to the leads \cite{Meir1991}. A strongly coupled ground state transition exhibits a 1/T-dependence for the peak height. In contrast, transport through a weakly connected ground state transition with a strongly coupled excitation is enhanced by activation. As $kT\gtrsim \Delta_s$ both levels contribute to transport.

We now focus on the thermal activation between resonances. In Fig. \ref{fig2}(c) we display the behavior at three representative back-gate values. In all cases shown here the conductance is temperature-independent at low $T$ and activation sets in for $T \gtrsim$ 3 K. The latter is linear in the logarithmic plot presented here, which is characteristic for activated transport. Thus, the data is fitted to the empirical law
\begin{equation}
G= G_\mathrm{0} \text{ exp}\left(-\frac{E_\mathrm{a}}{2k_\mathrm{B}T}\right) +G_\mathrm{off},
\label{G}
\end{equation}
where $G_\mathrm{0}$ is the large temperature conductance, $E_\mathrm{a}$ is the activation energy and $G_\mathrm{off}$ is a constant off-set. Eq. (\ref{G}) is used to fit the $T$-dependent conductance with those three parameters. It reproduces the data very well in all conductance valleys between resonances and even on some peaks.

\begin{figure}[ht]
  \begin{center}
    \includegraphics{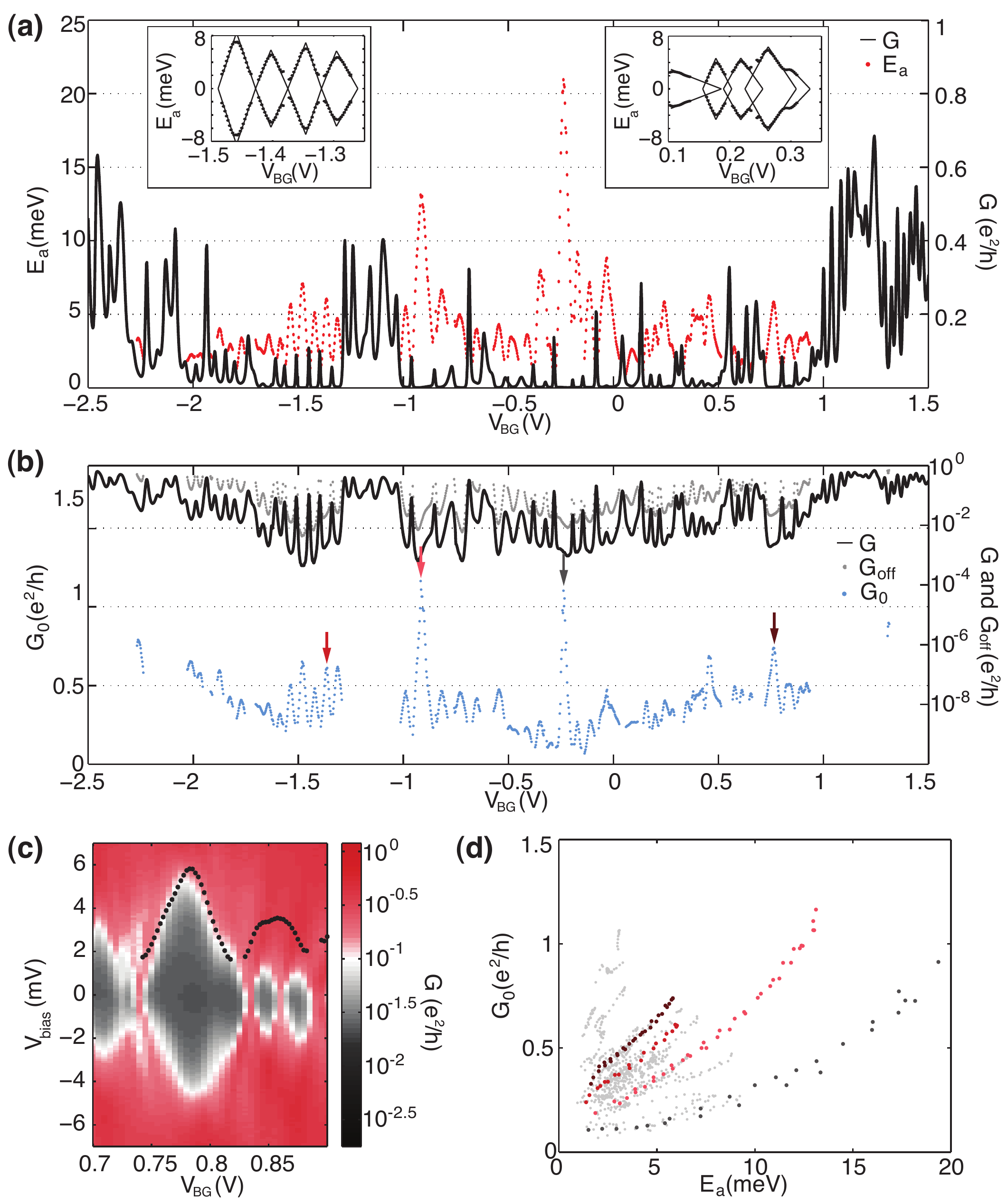}
    \caption{Back-gate dependence of fitting parameters: (a) $E_\mathrm{a}$, (b) $G_\mathrm{off}$ and $G_\mathrm{0}$. In (a) and (b) the black solid curve shows $G$ at base temperature. Insets: Coulomb diamonds reconstructed from $E_\mathrm{a}$ for two regimes. (c) Comparison of a Coulomb diamond (representing $E_\mathrm{c}(V_\mathrm{BG})$) and $E_\mathrm{a}$ determined for this $V_\mathrm{BG}$-interval. (d) $G_\mathrm{0}(E_\mathrm{a})$ for the transport gap. Colored branches indicate $G_\mathrm{0}$/$E_\mathrm{a}$ pairs that originate from the same conductance valleys (arrows in (b)).}
    \label{fig3}
  \end{center}
\end{figure}

With this model for transport in our system at hand we extract $E_\mathrm{a}$, $G_\mathrm{0}$ and $G_\mathrm{off}$ as a function of back-gate voltage. The analysis was performed only at those gate voltages where $G(T)$ spanned more than one order of magnitude. The results are shown in Figs. \ref{fig3}(a) and (b). Due to the given criterion for the analysis an evaluation at the edge of the transport gap as well as around $V_\mathrm{BG}$ = -1.25 V was not possible.

We start with discussing the high-temperature activated behavior found in the data. The activation energies peak in the middle between neighboring conductance resonances [Fig. \ref{fig3}(a)]. On the other hand, pronounced dips in the activation energies arise which coincide with conductance peaks . In-between a linear dependence on gate voltage is observed as is characteristic for Coulomb diamonds. Additionally, the largest $E_\mathrm{a}$ values are 10 to 20 meV. This energy scale is of the order of typical charging energies $E_\mathrm{c}$ of this device determined from the finite-bias spectroscopy in Fig. \ref{fig1}(c). As visualized in Fig. \ref{fig3}(c), a more careful comparison shows that the activation energy resembles the measured Coulomb diamond boundaries remarkably well. Due to thermal cycling in-between the diamond and the temperature measurement, some shifts are visible in the spectra if the two energy scales are plotted on top of each other over a large gate voltage range. The finding that the peak values of $E_\mathrm{a}$ in the valleys between conductance resonances are identical to the charging energy $E_\mathrm{c}$ extracted from Coulomb diamonds is a central result of this paper.

We can reconstruct Coulomb diamonds from the activation energy by mirroring $E_\mathrm{a}(V_\mathrm{BG})$ at the voltage axis and inserting lines along the linear slopes in $E_\mathrm{a}$. The insets of Fig. \ref{fig3}(a) displays two qualitatively different regions in back-gate voltage. In the left graph adjacent diamonds touch each other in one point at zero bias. Their size is similar and the flanks have the same slopes. For this back-gate voltage range the same observations are made for the boundaries of Coulomb blockade diamonds measured in finite-bias spectroscopy. Such a behavior is characteristic for a single quantum dot where levels are filled sequently. In the region under discussion, transport is therefore dominated by only one localized island. Since the charging happens from the (temperature broadened) leads, that are coupled to the island, the corresponding maximum $E_\mathrm{a}$ and $E_\mathrm{c}$ have to be interpreted as the on-site charging energy of this localized site. Its diameter corresponds roughly to the ribbon width when estimating the size of the puddle from $E_\mathrm{c}$ by a comparison to data taken on quantum dots.

The temperature dependence of the conductance resonances between these diamonds exhibits a monotonic increase [see Fig. \ref{fig2}(a)]. As discussed before, this is expected for multilevel transport \cite{Meir1991}.

As a second regime we chose a back-gate voltage in Fig. \ref{fig1}(c) around which the regions of suppressed current are connected to each other. The right inset in Fig. \ref{fig3}(a) shows the corresponding reconstruction of Coulomb diamonds from $E_\mathrm{a}$ where diamonds overlap and the size as well as the back-gate dependence of $E_\mathrm{a}$ vary strongly in neighboring diamonds. Taking this behavior as an indication for the participation of several dots in transport, we now have to attribute $E_\mathrm{a}$ and $E_\mathrm{c}$ to both on-site and inter-site charging energies. Stochastic Coulomb blockade describes such a phenomenon, where transmission through a small number of quantum dots is considered.

Next we proceed with a discussion of the low-temperature conductance represented by $G_\mathrm{off}$ in eq. (\ref{G}). We attribute $G_\mathrm{off}$,  which is evident in the curves in Fig. \ref{fig2}(c), to cotunneling processes that determine the conductance value before thermal activation sets in. Cotunneling leads essentially to Lorentzian tails of conductance resonances. The inset of Fig. \ref{fig2}(a) shows that indeed we can explain the resonance line shape taking into account both thermal and coupling broadening by a convolution of the derivative of the Fermi distribution with a Lorentzian. We can do a refined analysis of the low-temperature background by fitting the low $T$ data between resonances to the expression $G_\mathrm{low} \propto \beta (T^2+T_0^2)$ (not shown) \cite{Averin90}. Fig. \ref{fig3}(b) shows that the conductance spectrum taken at the lowest temperature is indeed reflected by the extracted cotunneling background. The finding of cotunneling transport supports the previous statement that only few islands are involved in transport since cotunneling becomes suppressed as the number of localized states increases. 

We now discuss the behavior of the prefactor $G_\mathrm{0}$ in Eq. (\ref{G}). It extrapolates the conductance for $k_\mathrm{B}T \gg E_\mathrm{a}$ and hence represents the high temperature conductance. The order of magnitude of $G_\mathrm{0}$ is between 0.1 and 1 in units of $e^2/h$. Similar to $E_\mathrm{a}$ it is strongly anti-correlated with the conductance at the lowest temperature as illustrated in Fig. \ref{fig3}(b). The correlation between $E_\mathrm{a}$ and $G_\mathrm{0}$ in conductance valleys is visualized in Fig. \ref{fig3}(d). Clearly, the $G_\mathrm{0}(E_\mathrm{a})$ plot consists of discrete branches with varying curvature/slope. Each color-coded branch corresponds to a peak of $E_\mathrm{a}$ in the back-gate spectrum. The ratio of $G_\mathrm{0}$ to $E_\mathrm{a}$ decreases as the pair originates from a back-gate value closer to the center of the transport gap.

Transport studies  in finite magnetic field have been carried out in two devices. As seen in earlier experiments \cite{Oostinga10} we find that the size of the Coulomb diamonds shrinks as a $B$-field is applied perpendicular to the graphene plane. This effect was attributed to time reversal symmetry breaking in the regime of strong localization which causes an increase of the conductance through the ribbon \cite{Shklovskii84,Gershenson97}. To get more evidence for the observed positive magnetoconductance we have investigated the temperature dependence at $B$ = 7 T. As for zero field the extracted maximum $E_\mathrm{a}$ is equal to $E_\mathrm{c}$ of the corresponding Coulomb diamond.

Comparing the different  ribbons under study we observe an increase of $\Delta V_\mathrm{BG}$ and $E_\mathrm{c}$ with decreasing ribbon width as discussed in other experiments \cite{Han07,Chen07,Liu09,Molitor09,Stampfer09,Todd09,Oostinga10,Han10,Gallagher10,Terres11,Tombros11}. The latter fact points to the formation of ever smaller islands, which block transport and lead to an increase of $E_\mathrm{c}$, as the ribbon gets narrower. The magnitude of the introduced energy $E_\mathrm{a}(V_\mathrm{BG})$ is tied to $E_\mathrm{c}(V_\mathrm{BG})$ for all measurements showing that they share the same physical origin.

Our temperature dependence differs from the one observed in Refs. \onlinecite{Han10} and \onlinecite{Oostinga10} where $G$ $\propto$ exp$(-T_\mathrm{0}/T)^{1/2}$ for low temperatures. Here, the large number of measured points in back-gate allowed us to analyze the temperature dependence for discrete $V_\mathrm{BG}$ values inside the transport gap. However, we can fit our data with the same temperature dependence as in Refs. \onlinecite{Han10} and \onlinecite{Oostinga10} if we apply the averaging methods used there.

The picture of transport we present here does not require but does not exclude either the contribution of phonons inside the system. Activation may take place in the leads from which the localized puddles get charged via smearing of the Fermi function. It is unclear whether phonons in the ribbon get important for transport at elevated temperatures. The origin of the correlation between $E_\mathrm{a}$ and $G_\mathrm{0}$ remains to be understood but may be linked to the role of phonons. 

In summary, we have studied thermally activated transport in graphene nanoribbons of different aspect ratios and compared the determined parameters to transport measurements at low temperature. We find that the transmission is dominated by mainly one of the few localized states inside the ribbon at a specific back-gate configuration. As a consequence transport in graphene nanoribbons should be understood as being mesoscopic and single particle-like and treated in such a framework.

We thank C. Beenakker, A. Morpurgo, J. Folk, F. Guinea and A. Yacoby for helpful discussions. This research was supported by the Swiss National Science Foundation through the National Centre of Competence in Research 'Quantum Science and Technology'.

\end{document}